\documentclass[superscriptaddress,twocolumn,preprintnumbers,nofootinbib,prl]{revtex4}
\usepackage{graphicx}

\usepackage[hypertex]{hyperref}

\newcommand{\lsim}{\lesssim}

\newcommand{\gsim}{\gtrsim}
\newcommand{\beq}{\begin{equation}}
\newcommand{\eeq}{\end{equation}}
\newcommand{\be}{B_\oplus}

\def\be{\begin{equation}}
\def\ee{\end{equation}}
\def\baray{\begin{eqnarray}}
\def\earay{\end{eqnarray}}
\def\ba{\begin{eqnarray}}
\def\ea{\end{eqnarray}}

\begin{document}

% page numbers bottom-center
\pagestyle{plain}

\preprint{BNL-HET-06/16}

\preprint{MAD-TH-06-10}

\preprint{MADPH-06-1469}

\preprint{CU-TP-1165}

\title{Vacuum Sampling in the Landscape during Inflation}
%\title{Cosmology and the Quantum Landscape}

\author{Hooman Davoudiasl
%\footnote{email: hooman@physics.wisc.edu}
}
\affiliation{Department of Physics, Brookhaven National
Laboratory, Upton, NY 11973-5000, USA}

\affiliation{Department of Physics, University of Wisconsin,
Madison, WI 53706, USA}

\author{Sash Sarangi}

\affiliation{Department of Physics, Columbia University, New York,
NY 10027, USA}

\author{Gary Shiu
%\footnote{email: shiu@physics.wisc.edu}
}

\affiliation{Department of Physics, University of Wisconsin,
Madison, WI 53706, USA}

%%%%%%%%%%%%%%%%%%%%%%%%%%%%%%%%%%%%%%%%%%%%%%%%%%%%%%%%%%%%%%%%%%%%%%%%%%%%

\begin{abstract}
We consider the phenomenological consequences of sampling multiple vacua 
during inflation motivated by an enormous landscape.   A generic consequence of this
sampling is the formation of domain walls, characterized by the
scale $\mu$ of the barriers that partition the accessed vacua. We
find that the success of Big Bang Nucleosynthesis (BBN) implies $\mu
\gsim 10$~TeV, as long as the sampled vacua have a non-degeneracy
larger than $\cal{O}({{\rm MeV}}^{\rm 4})$. Otherwise, the walls
will dominate and eventually form black holes that must reheat the
universe sufficiently for BBN to take place; in this case, we obtain
$\mu \gsim 10^{-5}M_P$.  These black holes are not allowed to
survive and contribute to cosmic dark matter density.

\end{abstract}
\maketitle

%%%%%%%%%%%%%%%%%%%%%%%%%%%%%%%%%%%%%%%%%%%%%%%%%%%%%%%%%%%%%%%%%%%%%%%%%%%

%\section{Introduction}

Various theoretical considerations have led to a picture of string
theory with a gigantic number of solutions, referred to as the
``Landscape" of vacua \cite{Susskind:2003kw}. This picture was
motivated in part by the construction of cosmological vacua in
Refs.~\cite{Kachru:2003aw, Kachru:2003sx}, and the dense
``discretuum'' in Refs.~\cite{Bousso:2000xa,Feng:2000if}. To
confront the vast landscape of string vacua, a statistical approach
has also been proposed \cite{Denef:2004ze}. Despite progress, a
detailed understanding of the topography of the landscape (such as
the distribution of minima and the heights of the barriers) remains
elusive.

Usually, effects due to
the multitude of vacua in the string
theory landscape
are relegated to the eternal inflationary phase in the
cosmological history of the universe. Once the universe enters the
last stage of inflation (that includes the last $60$ $e$-folds),
the multiple vacua cease to have any effect, and
the universe classically evolves towards a single vacuum in the
landscape. Portions of the universe in different vacua are
exponentially far away from each other and the landscape seemingly
does not leave any imprints on the post-inflationary cosmology.
Such scenarios have been considered, {\it e.g.}, in
Ref.~\cite{Linde:2005yw}.

Since inflation is a well-motivated ingredient of a consistent
cosmology and is strongly supported by data, we assume that it is
part of the cosmic history. The inflationary universe is generically
characterized by enormous energy scales, at which other vacua may be
accessed. The presence of an enormous landscape can then be expected
to affect cosmic evolution.  In this Letter, we consider the general
possibility that the fundamental description of Nature entails a
large landscape of vacua, some of which may be accessible during the
last $60$ e-folds of inflation. We then inquire how this premise
affects the cosmological evolution of the universe.

Inflationary vacuum-sampling may proceed in a variety of fashions.
For example, during inflation, characterized by a Hubble constant $H
$, a light field $S$ of mass $m_S \lsim H$ develops de Sitter
fluctuations $\delta S \sim H$ and may sample the local landscape.
In many situations, like in minimal models of supersymmteric
inflation, there are non-inflaton fields that are light during
inflation with masses a few orders of magnitude below $H$. Such
fields and their cosmological effects have been studied, for
example, in Refs.~\cite{Dvali:2003em, Linde:2005yw}.  If one starts
with an initial condition where $S$ is in one of the minima, given a
modest number of $e$-folds, the field can get delocalized and make
excursions to other vacua, for example because: {\it (i)} the
neighboring vacua are separated by barriers that are shallow or {\it
(ii)} due to resonance effects, tunneling transitions between
certain vacua take place unsuppressed. Such effects have recently
been utilized in the work of Ref.~\cite{Tye:2006tg}.

A demonstration of the relative likelihood of the above, or other,
mechanisms for vacuum sampling requires a detailed knowledge of the
fundamental description of how the landscape topography is generated 
microscopically and is not within the scope of
this work.  However, it has been argued that a large number ${\cal
N} \gsim 10^{120}$ of vacua may be necessary to accommodate a
Cosmological Constant (CC) of order $(10^{-30} M_P)^4$, as inferred
by the size of dark energy density. In fact, the success of
Weinberg's prediction \cite{Weinberg:1987dv} for the approximate
order of magnitude of the CC may point to a more dramatic
possibility that our local region of the landscape is a dense
attractor \cite{Dvali:2003br,Dvali:2004tm,Sakharov:1984ir}. Similar
attractor behavior also seem to appear in the landscape of flux
vacua \cite{Ashok:2003gk,Giryavets:2004zr,Conlon:2004ds}. Given such
a large number of vacua, one may expect that a few local directions
will be accessible to vacuum sampling during inflation. 
Given these considerations, we
take the possibility of more than one vacuum being accessed during
inflation to be a plausible assumption and will study its
phenomenological consequences.

%It is important that the coheron fluctuations do not
%back-react on the geometry strongly and we can treat such
%fields as perturbations on a fixed background.  To see this,
%note that the total energy density in the
%coheron field is of order $H^4$, whereas inflation is driven by
%an energy density $\rho_I \gg H^4$.

We expect vaccum sampling to be important only during  
inflation.  For example, if sampling is mediated by the 
above mentioned light scalar $S$, after inflation, this 
field no longer has de Sitter fluctuations and settles down
to one vacuum. Quite generally, we expect each causal patch of 
size $H^{-1}$ to settle into a different accessible
vacuum .  At the end of inflation and henceforth, we expect  
each patch will follow the usual single-vacuum evolution. 
However, different causal patches will be separated by domain walls.  
The future evolution will depend on whether the walls percolate or not.
This depends on the relative population of each vacua, i.e. the number of Hubble patches
that end up in different vacua. For two vacua, the relative population in each vacuum should
be above $0.31$ for percolation to happen. Presumably, the relative population will
depend on the split in the degeneracy $\epsilon^4$ between the vacua. For example, for the case
where the sampling of vacua happens due to Hubble fluctuations allowing the field to
hop over small barriers, the ratio of populations will go as $ \exp (- \epsilon/H )$.
Note that $\epsilon$ is bounded by the height of the barrier, which for sampling to work, should
be of the order $H$ or less. For $\epsilon$ an order of magnitude below $H$, the relative populations
easily satisfy the condition for percolation.  
If the walls percolate, the universe will consist of a long infinite wall of a complicated geometry
stretching across the $3$-dimensional space, and there will be a
distribution of finite clusters of walls. For $\epsilon$ close to $H$ percolation might not happen. 
If the walls do not percolate, then there will just be clusters
(bags of domain walls enclosing one type of vacuum within and the other type of vacuum outside) of various
sizes. The clusters should go away by the time of Big Bang Nucleosynthesis (BBN).  
%It seems plausible that if these patches do not percolate we cannot 
%expect to end up with a homogeneous and isotropic universe 
%we observe today.  In this case, vacuum sampling is disfavored, 
%strongly implying that the local landscape is dominated by 
%rather inaccessible vacua, even in the presence of 
%inflationary energy densities and quantum effects.  This conclusion 
%is fairly non-trivial, since the physics of inflation is generically 
%close to the string scale.

From now on we consider the case in which the domains percolate and 
horizon-sized domain walls form.  
The tension of these domain walls depends on the moduli, set by the
local topography of the landscape. These walls can in
principle lead to serious problems for cosmology
\cite{Zeldovich:1974uw}. First of all, they can quickly come to
dominate the energy density of the universe and disrupt radiation
domination that must set in, say, by the time of the BBN.  Secondly, if the walls exist after
recombination, unless their tension is below $({\rm MeV})^3$, 
they would lead to large 
anisotropies in the cosmic microwave background radiation.

Therefore, it seems that our picture has led to an
unacceptable state for the early universe.  However, this is not
necessarily the case.  Domain walls are only stable if they come
from spontaneously broken discrete symmetries.  In that case, the
various domains are degenerate vacua of the theory.  During vacuum 
sampling of the landscape, however, there is no symmetry
that would require different causal
patches to have the same vacuum energy. If the vacua are not
degenerate, as is well-known \cite{Gelmini:1988sf,Preskill:1991kd},
the unwanted walls can be pushed away into the false vacuum regions, by the
pressure from the energetic bias, resulting in a universe
dominated by the true vacuum. We will next study the phenomenological
constraints that these considerations may impose on the local landscape.

The above solution to the domain wall problem is usually based on
a scenario in which the universe is partitioned into two phases by
a typically complicated domain wall.  The two-phase universe is a
result of percolation, a process that depends on the relative
probability for each phase to emerge in a given causal patch.  The
situation resulting from vacuum sampling of the landscape could,
in principle, be more complicated.  It is a possibility that a
sufficient number of local minima might be accessible so that the
domain walls are approximately (as these symmetries are
not exact) of the type $Z_N$, with $N > 2$. $Z_N$ domain walls
typically consist of configurations of $N$ domain walls attached
to a string \cite{VS}.

To have a more transparent analysis of the domain wall problem, let
us consider the simple case of only two accessible vacua at the end
of inflation.  This corresponds to having two dominant vacuum populations, 
one of which leads to our universe.  Our treatment follows
considerations similar to those in Ref.~\cite{Vilenkin:1981zs}.
The wall formation happens at the end of inflation (as access to 
multiple vacua is lost), given by the time scale
\beq
t_{wf} \simeq t_I \sim 1/H,
\label{twf}
\eeq
where the subscript $I$ denotes the end of
inflation. Here, $H \sim M_I^2/M_P$, where $M_I$ is the is the
inflationary scale, and the $M_P$ is the 4-$d$ Planck scale.  We
will assume that the wall is flat over a causal patch.  Thus, the
energy density in the wall, characterized by a scale $\mu$,
is given by $\rho_w \sim \mu^3/t$.
$t$ is  given by the size of the causal patch.  The radiation energy
density is then given by $\rho_r \sim M_P^2/t^2$. We take the
universe to be initially radiation
dominated, which is a typical assumption \cite{Gelmini:1988sf}.
Domain wall
domination ($\rho_w > \rho_r$) begins at a time scale \beq t_{wd}
\sim M_P^2/\mu^3. \label{twd} \eeq Until time $t_{wd}$, the
evolution of the post-inflationary universe is governed by radiation
domination; $a \propto t^{1/2}$, where the scale factor is denoted
by $a$.

Once the domain walls form, they can lead to two alternative
cosmologically viable scenarios. The first scenario requires that
vacuum states separated by walls do not have degenerate energy
densities. As we discussed before, if $\epsilon \sim H$, percolation
might not happen, and then, as we shall show, the clusters of domain walls will disappear
automatically. We now consider the case where $\epsilon$ is small
compared to $H$ such that percolation does happen.
As a toy model of the landscape we consider just two
vacua with an energy density difference $\epsilon ^4$. Here, we
require that the walls disappear before they dominate the
universe. This can happen if $\epsilon > \epsilon_{*}$ which we
shall now determine. The bias $\epsilon$ becomes important at time
$t_\epsilon$ when the volume energy in the domain wall is of the order of
the surface energy:
\beq
t_\epsilon \sim \mu^3/\epsilon^4.
\label{teps}
\eeq
This energy bias will make the lowest energy
vacuum prevail and make the domain walls disappear.
Requiring $t_\epsilon < t_{wd}$, we get
\beq
\epsilon_{*}^4 \sim \mu^6/M_P^2
\eeq
However, we
still do not want the walls to disrupt well-established
cosmological processes.  Thus, we conservatively require that the
walls disappear before BBN: $t_{wd} < t_{BBN}$. This yields $\mu
\gsim 10$~TeV ; $\epsilon^4_* \gsim $ MeV$^4$.

If $\epsilon < \epsilon_*$, we have the second scenario. The
universe becomes domain wall dominated at $t \sim t_{wd}$ and the
domain walls grow till there is a gravitational instability that
makes them collapse into black holes. In fact, the time at which
the wall collapses to black holes coincides with $t_{wd}$,
i.e. $t_{collapse} = t_{wd}$. To see this let us first
estimate the mass of a
black hole thus formed. Consider a sphere of radius $R$ centered
on the wall. The mass of the wall inside the sphere is given by
$M(R) \sim R^2 \mu^3$. (We assume that the energy due to the bias
$\epsilon$ at such radius $R$ is less than that due to the wall
tension.  This is a justified assumption since $\epsilon <
\epsilon_{*}$ and the collapse
happens before the volume term dominates.)
The radius grows and when $R = R_{bh}\sim
M_P^2/\mu^3$ the condition $M_P^{-2} M(R) = R$ will be satisfied
and the wall will collapse into a black hole. The mass of the black
hole is
\beq
M_{bh} \sim (M_P/\mu)^3 M_P
\eeq
Now if we use the time dependence of $R$, $R \sim t$, then we see that
the time at which the wall collapses to black holes is given by
$t_{collapse} \sim M_P^2 /\mu^3$, coinciding with $t_{wd}$. When the wall
is about to become the dominant form of energy, it collapses into black holes.
The universe will now
become black hole dominated. However depending on how light these
black holes are, they can quickly evaporate. The time scale for
black hole evaporation is
\beq
\tau_{bh} \sim M_{bh}^3/M_P^4.
\label{tau}
\eeq
If the
wall tension is set by $\mu \sim 10^{-3} M_P$, which can be a
reasonable string scale $M_s$, then we see that black holes will
evaporate within $\tau_{bh} \sim 10^{-16}$~s.

An interesting outcome of the above process is that black hole
evaporation can {\it reheat} the universe to high temperatures
again.  Thus, vacuum sampling may lead to
{\it landscape-assisted} reheating.
To see this, we note that at formation, the energy
density in black holes is 
\beq
\rho_{bh}(t_{wd}) \sim \mu^6/M_P^2.
\label{rhotwd}
\eeq
After the collapse of the walls into black holes
at $t_{wd}$, the evolution of the scale factor is
that of a matter dominated universe.  Therefore, at the time
of black hole evaporation $\tau_{bh}$, the scale factor is given
by
\beq
\frac{a_\tau}{a_{wd}}=\left(\frac{\tau_{bh}}{t_{wd}}\right)^{2/3}
\!\!\sim  \left(\frac{M_P}{\mu}\right)^4.
\label{aevol}
\eeq

Thus, the energy density locked up in black holes just before
their decay, which we approximate as an instantaneous event,
is
\beq
\rho_{bh}(\tau_{bh}) \sim \mu^{18}/M_P^{14}.
\label{rhotaubh}
\eeq
If we assume that the above energy density is converted
into radiation at $t \sim \tau_{bh}$, then the resulting
temperature of the universe is estimated by 
%$T_{rh} \sim [\rho_{bh}(\tau_{bh})]^{1/4}$.
\beq
T_{rh} \sim [\rho_{bh}(\tau_{bh})]^{1/4}.
\label{Trh}
\eeq
For $\mu = 10^{-3} M_P$, as before, we get $T_{rh} \sim
100$~TeV, far above the weak scale,
which is an acceptable reheat temperature.  $T_{rh}$
should be at least of order 1~MeV to allow for the BBN.
Hence, we conclude that to exit black hole domination
gracefully, we must demand $\mu > \mu_* \sim 10^{-5} M_P$.
Even if reheating can be acheived by the inflaton instead of
black holes, for $\mu$ below this bound, the universe
would not have exited the black hole domination phase
quickly enough to allow for standard BBN.

One could envisage a situation where there is a distribution
in the masses of the black holes such that some of them
might survive till today.
These black holes essentially provide bridges
to neighboring minima of the
landscape while they exist.
The spectrum of the masses of the black holes will
depend on the detailed structure of the local string landscape.
Depending on the spread in the heights of the barriers between
neighboring minima, there will be a spread in black hole
masses. However, as we shall see now, even 
with a mass distribution, all black holes 
should evaporate by reheating. That is to say, the heaviest 
black hole population (resulting from the collapse of the walls with the least tension) 
must be responsible for post-inflationary reheating.

To demonstrate this point, let us consider a bimodal distribution of wall
tensions with walls of type A with mass scale $\mu_A$ and wall
of type B with tension $\mu_B$. We would like wall A to reheat
the universe via the evaporation of the black holes it would
collapse to. So the previous analysis regarding black hole
assisted reheat applies to wall A. The reheat energy is given by
$\rho_{reheat} \sim \mu_A^{18}/M_P^{14}$. Now we would like to
see if the black holes formed due to the collapse of the
wall B can be entirely responsible for dark matter. For these
black holes to have survived till today, we must have $\mu_B
<< \mu_A$. This justifies the neglect of wall B when we
derive the reheat scale. The time of collapse
of wall B into black hole B be $t_{collapse}^{B} \sim M_P^2/\mu_B^3$.
However, just as for wall A, the $t_{collapse}^B$ will coincide
with the time when wall B begins to be the dominant form of
energy in the universe. 

To see this, one just has to realize
that at all stages of the cosmological evolution, the energy
density is redshifting as $1/t^2$. First there is a radiation
dominated stage, followed by a black hole A dominated stage,
followed by another radiation dominated stage due to reheat,
and then at some time wall B begins to dominate. The condition for
wall B domination is still $M_P^2/t^2 \sim \mu_B^3/t$, i.e.
$t_{wd}^B \sim M_P^2/\mu_B^3$. 

Hence, the time at which the
energy density begins to be the dominant energy contribution
exactly coincides with the time at which wall B collapses to
black hole B and becomes the dominant component. Since we
would like to use black hole B as dark matter, this time
must be the time of radiation-matter equality, at $T_{eq}\sim 1$~eV: 
\beq
t_{wd}^B \sim t_{collapse}^B \sim M_P^2/\mu_B^3 \sim M_P/T_{eq}^2.
\label{teq}
\eeq
This fixes the value of
the mass scale of wall B, $\mu_B \sim 1$~GeV. The mass of the resulting
black holes is $\left( M_P/\mu_B \right)^3 M_P \sim 10^{19} M_\odot$, where 
$M_\odot \sim 10^{30}$~kg is the solar mass.  
Such black holes are too heavy to make dark matter, since a typical galactic 
mass is of order $10^{11} M_\odot$.

Incidentally, this leads to a no-go theorem for walls that form around
inflation to give dark matter via collapse into black holes. Let us
consider $k$ different black holes with different tensions. We would like
the last surviving black holes to give dark matter. All the previous
black holes just release energy into radiation as they evaporate.
However, as we have seen above, the time of the collapse of a wall
into black holes exactly coincides with the time when the wall begins
to be the dominant form of energy. This time is given by $t \sim M_P^2/\mu_k^3$.
However since we want these blackholes to be dark matter, 
$t \sim M_P/T_{eq}^2$, the
time of radiation-matter equality. This fixes the mass of the black holes
to be $10^{19} M_\odot$. Hence, we have a no-go theorem: {\it Walls that form
from inflationary vacuum sampling cannot yield black hole dark matter.}

The minimum bias
$\epsilon_* \sim$~MeV we have derived suggests that unless
the sampled vacua have large vacuum energies compared to that in our
observable universe, we end up in a black hole dominated
universe.  In this case, the walls must be predominantly
characterized by a scale
larger than $\mu_* \sim 10^{-5} M_P$.
Put another way, either none of the vacua in
our local landscape are accessible to inflationary
sampling, or else all the accessible ones have large
cosmological constants or high barriers.
The above bounds are at best rough
estimates and subject to well defined assumptions.  However,
because the basic ingredients
leading to these conclusions are general relativity, inflation,
and quantum mechanics, the above considerations point to possible
bottom up constraints on the features of the local landscape.
It is to be noted that even in the first scenario, where
wall domination is avoided by having a bias in the vacuum energies
of the vacua, we derived a lower bound on the wall tension 
$\mu \gsim 10$~TeV 
by requiring that the walls be gone before BBN, as a minimal normalcy 
condition. 

In summary, we assumed that one or
more efficient ways of transit to
other landscape minima exist and access to a few
vacua during inflation may well be plausible.  This 
assumption is motivated by the large number of possible 
neighboring vacua and the closeness of inflationary and  
string scales.  We examined domain wall formation, as a
generic consequence of vacuum
sampling during inflation.  The fasle vacua can 
be pushed out given a bias between the sampled vacuum energies.  BBN 
considerations point to a minimum bias of order MeV$^4$.  The barriers 
are then characterized by a minimum scale of order 10~TeV. 
In the absence of the necessary vacuum energy bias, domain-wall domination 
will ensue and lead to the genesis of
primordial blackholes.  The evaporation of these
balck holes must reheat the universe.  BBN constraints then suggest 
a minimum barrier scale of order $10^{-5}M_P$.  We showed that these 
priomordial balck holes cannot survive to be part of cosmic dark matter.  
Our bottom-up approach could be viewed as a phenomenological 
guide for future landscape model-building.

We thank X.~Chen, P.~Corasaniti, B.~Greene, D. Kabat,
J.~Polchinski, and H.~Tye for discussions.
SS would like to thank the particle theory group
at University of Wisconsin, Madison, and the Perimeter Institute.
This work was supported in part by the DOE
under contracts DE-AC02-98-CH-10886 (HD),
DE-FG02-95ER40896 (HD, GS) and DE-FG02-92ER40699 (SS),
NSF CAREER Award PHY-0348093 (GS), a Research Innovation
Award (GS) and a Cottrell Scholar Award (GS) from Research Corporation,
and the P.A.M. Dirac Fellowship (HD)
at the University of Wisconsin-Madison.
%%%%%%%%%%%%%%%%%%%%%%%%%%%%%%%%%%%%%%%%%%%%%%%%%%%%%%%%%%%%%%%%%%%%%%%%%%%

\end{document}